\documentclass[pra,showpacs,twocolumn,aps,superscriptaddress,a4paper]{revtex4-1}
\usepackage{dcolumn,amssymb,amsmath,amsfonts,graphicx,latexsym,color}
\usepackage{epstopdf}
\usepackage{CJK}
\begin{document}

\begin{CJK*}{GBK}{song}

\title{Universal relations and normal phase of an ultracold Fermi gas with coexisting $s$- and $p$-wave interactions}
\author{Fang Qin}
\email{qinfang@ustc.edu.cn}
\affiliation{Key Laboratory of Quantum Information, University of Science and Technology of China, Chinese Academy of Sciences, Hefei, Anhui 230026, China}
\affiliation{Synergetic Innovation Center of Quantum Information and Quantum Physics, University of Science and Technology of China, Hefei, Anhui 230026, China}
\author{Xiaoling Cui}
\email{xlcui@iphy.ac.cn}
\affiliation{Beijing National Laboratory for Condensed Matter Physics, Institute of Physics, Chinese Academy of Sciences, Beijing 100190, China}
\author{Wei Yi}
\email{wyiz@ustc.edu.cn}
\affiliation{Key Laboratory of Quantum Information, University of Science and Technology of China, Chinese Academy of Sciences, Hefei, Anhui 230026, China}
\affiliation{Synergetic Innovation Center of Quantum Information and Quantum Physics, University of Science and Technology of China, Hefei, Anhui 230026, China}

\date{\today}

\begin{abstract}
We study the universal relations and normal-phase thermodynamics of a two-component ultracold Fermi gas with coexisting $s$- and $p$-wave interactions. Due to the orthogonality of two-body wave functions of different scattering channels, the universal thermodynamic relations of the system appear to be direct summations of contributions from each partial-wave scattering channels. These universal relations are dictated by a set of contacts, which can be associated with either $s$- or $p$-wave interactions. Interestingly, due to the interplay of $s$- and $p$-wave interactions on the many-body level, the contacts, and hence all the relevant thermodynamic quantities, behave differently from those with only $s$- or $p$-wave interactions. These are manifest in our numerical calculations based on second-order virial expansions for $^{40}$K atoms under typical experimental parameters. A particularly interesting finding is that, due to the coexistence of $s$- and $p$-wave scatterings, the interaction energy of the repulsive branch features abrupt changes across the $p$-wave resonances. Our results can be readily checked experimentally for $^{40}$K atoms near the $198$G $p$-wave Feshbach resonance, where multiple partial-wave scatterings naturally coexist.
\end{abstract}


\maketitle

\section{Introduction}

In ultracold atoms, the typical diluteness condition and short-range interactions can give rise to interesting universal relations among thermodynamic quantities of a many-body system. These universal relations are independent of the microscopic details of two-body interactions, and have stimulated much research interest. In a unitary Fermi gas close to an $s$-wave scattering, for instance, the central quantity of these universal relations, Tan's contact, has been extensively studied both theoretically and experimentally. Furthermore, the recent radio-frequency (r.f.) spectroscopic measurement near a $p$-wave Feshbach resonance in $^{40}$K atoms opens up new possibilities of studying universal relations and thermodynamics in higher partial-wave scattering channels~\cite{exp2004s,exp2003p,review2010,exp2004p,Yu2015exp}. Theoretically, it has also been shown recently that a whole family of contacts and universal relations exist in systems with higher partial-wave scattering~\cite{Zhang2009,Castin20121,Castin20122,Zwerger2016,Yu2015,Peng2016,Yoshida2015,Zhou2016,Yu2016,Zhou20162,Yoshida2016,Cui20161,Cui20162,Qi2016,Tan2013,Tan2014,Peng2015}.

Motivated by this progress, we investigate universal relations and thermodynamics of a two-component degenerate Fermi gas with coexisting $s$- and $p$-wave interactions. Such a system can be experimentally prepared and probed with $^{40}$K atoms near the magnetic field $198$G. As the $p$-wave Feshbach resonance here is also close to an $s$-wave Feshbach resonance, the $s$- and $p$-wave interactions can be comparable in strength. It is then expected that the co-existence of both $s$- and $p$-wave scattering channels can lead to interesting many-body properties. Indeed, it has been shown very recently that, at zero temperature, an interesting hybridization of $s$- and $p$-wave superfluid can be stabilized, which exhibit nontrivial pairing correlations~\cite{Zhou2015sp}. In this work, we focus on thermodynamic properties of the normal state above the superfluid transition temperature.

We first derive the general universal relations of a two-component ultracold Fermi gas with coexisting $s$- and $p$-wave interactions. Due to the orthogonality of two-body wave functions of different scattering channels, the universal thermodynamic relations of the system, such as the adiabatic relations, pressure, and the energy functional, appear to be direct summations of contributions from $s$- and $p$-wave interactions. These universal relations are governed by a set of contacts, which can be associated with scatterings within different partial-wave channels. We then numerically evaluate the interaction energy as well as contacts in the high-temperature normal phase using second-order virial expansions. Interestingly, due to the interplay of $s$- and $p$-wave interactions on the many-body level, we find that the contacts, and hence the relevant thermodynamic properties  behave differently from those with only $s$- or $p$-wave interaction potentials. The interplay of multiple partial-wave interactions also leaves experimentally detectable signatures in asymptotic behavior of momentum distribution and r.f. spectrum in the high-momentum, high-frequency regime.
Another interesting finding is that the interaction energy of the repulsive branch features abrupt changes across the $p$-wave resonances, which may serve as a unique signature for the coexistence of $s$- and $p$-wave interactions in the system. Our results can be readily checked experimentally for $^{40}$K atoms near the $198$G $p$-wave Feshbach resonance, where the two scattering channels naturally coexist.

The paper is organized as follows: In Sec.~\ref{2}, starting from the two-body density matrix, we derive the general universal relations such as interaction energy, adiabatic relations, pressure relations, and the energy functional. In Sec.~\ref{3}, we present the formalism of the quantum virial theorem, and express thermodynamic quantities such as the thermodynamic potential and interaction energy in the normal state using second virial coefficients. In Sec.~\ref{4}, we numerically evaluate interaction energy and contacts using virial expansion calculations, adopting typical experimental parameters of $^{40}$K atoms near the $198$G $p$-wave Feshbach resonance. We also calculate the asymptotic behavior of the momentum distribution and the r.f. spectrum in the large-momentum and large-frequency limit, respectively. Finally, we summarize in Sec.~\ref{5}.

\section{Universal relations}\label{2}

In this section, we study the universal relations for a two-component, partially polarized Fermi gas with both $s$- and $p$-wave interactions. Following the standard treatment, we start from the two-body density matrix of the many-body system, decompose it into pair wave functions, and expand the pair wave functions in terms of the two-body wave functions. Employing the asymptotic forms of the two-body wave functions in the regime $1/k\gg r\gg r_0$, where $r_0$ is the range of interaction potentials, we derive universal thermodynamic relations based on the two-body density matrices. Importantly, these universal relations are governed by a set of contacts, which are associated, respectively, with $s$- and $p$-wave scattering channels. For the derivations in this section, we consider the general case where $s$-wave interactions between the two spin components ($|\uparrow\rangle, |\downarrow\rangle$) and $p$-wave interactions between the same spins coexist.

\subsection{Interaction energy and contacts}\label{2.2}

Following the standard treatment, we consider the two-body density matrix for our many-body system $\rho_2^{(\eta\sigma\gamma\delta)}(\vec{r}_1,\vec{r}_2)\equiv \langle\psi_{\sigma}^\dagger(\vec{r}_1)\psi_{\eta}^\dagger(\vec{r}_2)\psi_{\gamma}(\vec{r}_2)\psi_{\delta}(\vec{r}_1)\rangle$, where $\sigma,\gamma, \eta, \delta$ are spin indices, and $\psi_{\sigma}^\dagger(\vec{r})$ creates a fermion with spin $\sigma$ at position $\vec{r}$. The two-body density matrix can be decomposed as~\cite{Zhang2009,Yu2015,Leggett2006}
\begin{eqnarray}
\rho_2^{(\eta\sigma\gamma\delta)}(\vec{r}_1,\vec{r}_2)=\sum_{\alpha} n^{(\eta\sigma\gamma\delta)}_{\alpha}\phi^{(\eta\sigma)*}_{\alpha}(\vec{r}_1,\vec{r}_2)\phi^{(\gamma\delta)}_{\alpha}(\vec{r}_1,\vec{r}_2),\nonumber\\
\end{eqnarray}
where $\alpha=\{\vec{P}_c,j,l,m\}$, $\vec{P}_c$ is the center-of-mass momentum of a pair, $n^{(\eta\sigma\gamma\delta)}_{\alpha}$ are the eigenvalues, and $j,l,m$ label the quantum numbers of the relative radial direction, the angular momentum, and the $z$-axis projection of angular momentum, respectively. Moreover, the associated pair wave functions $\{\phi^{(\eta\sigma)}_{\alpha}(\vec{r}_1,\vec{r}_2)\}$ form an orthonormal set.

For the system under consideration, the only nonvanishing pair wave functions are
\begin{eqnarray}
\phi^{(\uparrow\uparrow)}_{\vec{P}_c,j,1,m}(\vec{r}_1,\vec{r}_2)&&=\tilde{\chi}_{11}\tilde{\phi}_{\vec{P}_c,j,1,m}^{(\uparrow\uparrow)}(\vec{r}_1,\vec{r}_2),\\
\phi^{(\downarrow\downarrow)}_{\vec{P}_c,j,1,m}(\vec{r}_1,\vec{r}_2)&&=\tilde{\chi}_{1-1}\tilde{\phi}_{\vec{P}_c,j,1,m}^{(\downarrow\downarrow)}(\vec{r}_1,\vec{r}_2),\\
\phi^{(\uparrow\downarrow)}_{\vec{P}_c,j,0,0}(\vec{r}_1,\vec{r}_2)&&=\tilde{\chi}_{00}\tilde{\phi}_{\vec{P}_c,j,0,0}(\vec{r}_1,\vec{r}_2),
\end{eqnarray}
with $m=-1,0,1$. The spin wave functions are defined as
\begin{eqnarray}
&&\tilde{\chi}_{00}=\frac{\sqrt{2}}{2}(|\uparrow\rangle_1|\downarrow\rangle_2-|\downarrow\rangle_1|\uparrow\rangle_2),\\
&&\tilde{\chi}_{11}=|\uparrow\rangle_1|\uparrow\rangle_2,\\
&&\tilde{\chi}_{1-1}=|\downarrow\rangle_1|\downarrow\rangle_2,
\end{eqnarray}
and the spatial wave functions
\begin{eqnarray}
&&\tilde{\phi}_{\vec{P}_c,j,0,0}(\vec{r}_1,\vec{r}_2)=\frac{\exp(i\vec{P}_{c}\cdot\vec{R}_{c})}{\sqrt{V}r}\varphi_{j,0,0}(r)Y_{0,0}(\hat{r}),\\
&&\tilde{\phi}_{\vec{P}_c,j,1,m}^{(\sigma\sigma)}(\vec{r}_1,\vec{r}_2)=\frac{\exp(i\vec{P}_{c}\cdot\vec{R}_{c})}{\sqrt{V}r}\varphi_{j,1,m}^{(\sigma\sigma)}(r)Y_{1,m}(\hat{r}).
\end{eqnarray}
Here, $\vec{R}_{c}=(\vec{r}_1+\vec{r}_2)/2$, $\vec{r}=\vec{r}_1-\vec{r}_2$, $V$ is the quantization volume, and
$Y_{l,m}(\hat{r})$ is the relevant spherical harmonic function. Correspondingly, we denote the relevant eigenvalues as $n^{(\uparrow\downarrow)}_{\vec{P}_{c},j,0,0}$ and $n^{(\sigma)}_{\vec{P}_{c},j,1,m}$, for the $s$- and $p$-wave pair wave functions, respectively.


We may then write down the interaction energy of the many-body system in terms of the two-body density matrix, and expand the pair wave functions therein using two-body radial wave functions. The resulting interaction energy reads
\begin{eqnarray}\label{U}
\langle \mathcal{U}\rangle
&&=\frac{1}{4\pi}C_s\int dr~U_s(r)|\chi_{s}(r)|^2 \nonumber\\
&&~~ +  \frac{1}{8\pi} \sum_{m,\sigma=\uparrow,\downarrow}\left[C_{p,\nu_{m}}^{(\sigma\sigma)}\int dr~U_p^{(\sigma\sigma)}(r)|\chi^{(\sigma\sigma,0)}_{p,m}(r)|^2\right. \nonumber\\
&& \left.~~+2C_{p,R_{m}}^{(\sigma\sigma)}\int dr~U_p^{(\sigma\sigma)}(r)\chi^{(\sigma\sigma,0)}_{p,m}(r)\chi^{(\sigma\sigma,1)}_{p,m}(r)\right].\nonumber\\
\end{eqnarray}
Here, we have defined a set of contacts $C_{s}$ and $C_{\nu_{m}/R_{m}}^{(\sigma\sigma)}$, which are related to the scattering length $a_s$, the scattering volume $\nu_{m}^{(\sigma\sigma)}$, and effective range $R_{m}^{(\sigma\sigma)}$, respectively
\begin{eqnarray}
&&C_s =4\pi\sum_{\vec{P}_{c},j}n^{(\uparrow\downarrow)}_{\vec{P}_{c},j,0,0}\int dk~a_{j,k}\int dk'~a_{j,k'},\\
&&C_{p,\nu_{m}}^{(\sigma\sigma)} =4\pi\sum_{\vec{P}_{c},j}n^{(\sigma)}_{\vec{P}_{c},j,1,m}\int dk~b_{j,m,k}^{(\sigma\sigma)}\int dk'~b_{j,m,k'}^{(\sigma\sigma)},\nonumber\\
\\
&&C_{p,R_{m}}^{(\sigma\sigma)} =2\pi\sum_{\vec{P}_{c},j}n^{(\sigma)}_{\vec{P}_{c},j,1,m}\nonumber\\
&&~~\times\int dk\int dk'~b_{j,m,k}^{(\sigma\sigma)}b_{j,m,k'}^{(\sigma\sigma)}(k^2+k'^2).
\end{eqnarray}
Note that the integration over $k$ here includes the contributions of the bound states. Note that the orthogonality of the two-body wave functions in different scattering channels renders the interaction energy in Eq.~\eqref{U} a direct summation of contributions from the three scattering channels: one with the $s$-wave interaction between $|\uparrow\rangle$ and $|\downarrow\rangle$, and two with the $p$-wave interaction between two $|\uparrow\rangle$s or two $|\downarrow\rangle$s. As a result, different scattering channels are decoupled in the definition of contacts as well. As we will see, this effectively allows us to write down the universal relations based on an intuitive extension of the previous works on single-partial-wave interactions.

\subsection{Adiabatic relations}\label{2.3}

The importance of contacts is that they appear in various universal relations and provide a link between few-body parameters and many-body quantities. This link is most apparent in the so-called adiabatic relations. Following the treatment in Refs.~\cite{Zhang2009,Yu2015}, we have
\begin{eqnarray}
\label{Cs1}&&\left.\frac{dE}{da_s^{-1}}\right|_{\nu_{m},R_{m}}=-\frac{\hbar^2}{4\pi M}C_s,\\
\label{CpV1}&&\left.\frac{dE}{d(\nu_{m}^{(\sigma\sigma)})^{-1}}\right|_{R_{m},a_s} =  -\frac{\hbar^2}{8\pi M} \sum_{m} C_{p,\nu_{m}}^{(\sigma\sigma)},\\
\label{CpR1}&&\left.\frac{dE}{d(R_{m}^{(\sigma\sigma)})^{-1}}\right|_{\nu_{m},a_s} = -\frac{\hbar^2}{8\pi M} \sum_{m} C_{p,R_{m}}^{(\sigma\sigma)},
\end{eqnarray}
where $E$ is the total energy, $\hbar$ is the reduced Planck constant, and $M$ is the atomic mass. Again, in the universal relations, we see that contacts associated with different scattering channels are decoupled.

\subsection{Pressure relation and virial theorem }\label{2.4}

Based on the adiabatic relations, it is straightforward to derive other related universal relations. As two illustrating examples, we show below the pressure and the energy functional.

The pressure for a uniform gas is given by~\cite{Zhang2009,Yu2015,Tan20081,Tan20082,Tan20083}
\begin{eqnarray}\label{Ph}
\mathcal{P}&&=\frac{2}{3}\frac{E}{V}+\frac{\hbar^2}{4\pi MV}\left[\frac{C_s}{3 a_{s}}\right. \nonumber\\
&& \left.~~ + \frac{1}{2} \sum_{m,\sigma=\uparrow,\downarrow} \frac{C_{p,\nu_{m}}^{(\sigma\sigma)}}{\nu_{m}^{(\sigma\sigma)}} +  \frac{1}{6} \sum_{m,\sigma=\uparrow,\downarrow}  \frac{C_{p,R_{m}}^{(\sigma\sigma)}}{R_{m}^{(\sigma\sigma)}}\right].
\end{eqnarray}
We note that although Eq.~(\ref{Ph}) is only exact in the absence of external potential, it is approximately valid for each local part of a trapped Fermi gas within the local density approximation. In an external harmonic trap $V_{ext}(\vec{r})=M\omega_{os}^{2}r^{2}/2$, where $\omega_{os}$ is the frequency of the harmonic trapping potential, the average energy density is replaced by the local internal energy density, and the contacts are replaced by the local contacts.

Furthermore, the total energy in the presence of a trapping potential is given by~\cite{Zhang2009,Castin20121,Yu2015,Tan20083,Braaten2008,Braaten20082,Braaten2011,Werner2008}
\begin{eqnarray}
E&&= 2\langle\mathcal{V}\rangle - \frac{\hbar^2}{8\pi M}\left[\frac{C_s}{a_{s}}\right. \nonumber\\
&& \left.~~ + \sum_{m,\sigma=\uparrow,\downarrow} \frac{3C_{p,\nu_{m}}^{(\sigma\sigma)}}{ 2\nu_{m}^{(\sigma\sigma)}}+\sum_{m,\sigma=\uparrow,\downarrow }\frac{C_{p,R_{m}}^{(\sigma\sigma)}}{ 2R_{m}^{(\sigma\sigma)}}\right],
\end{eqnarray}
where $\langle\mathcal{V}\rangle$ is the energy of the trapping potential. The relation above is essentially the virial theorem in Refs.~\cite{Zhang2009,Yu2015,Tan20083}.

\section{Virial expansion for a spin polarized Fermi gas}\label{3}

To further study the thermodynamic properties of the system, we evaluate the thermodynamic potential using virial expansions. For a spin-polarized Fermi gas, it is necessary to introduce two fugacities $z_{\uparrow} \equiv \text{exp}( \beta\mu_{\uparrow} )$ and $z_{\downarrow} \equiv \text{exp}( \beta\mu_{\downarrow} )$ to distinguish different spin configurations, where $\beta=1/(k_B T)$, $\mu_{\sigma}$ is the chemical potential for spin $\sigma$, $T$ is the temperature, and $k_B$ is the Boltzmann constant. Generally, the thermodynamic potential can be written as~\cite{liu3Dimbalance,liureview2013}
\begin{eqnarray}
\Omega(\mu_{\uparrow},\mu_{\downarrow})
= -\frac{1}{\beta}Q_{1}\sum_{n=1}^{\infty}\sum_{j=0}^{n}z_{\uparrow}^{n-j}z_{\downarrow}^{j}b_{n,j},
\end{eqnarray}
where $b_{n,j}$ is the $n$th virial coefficient associated with a cluster containing $n-j$ spin-$\uparrow$ fermions and $j$ spin-$\downarrow$ fermions. Here, $Q_{1}=2V/\lambda^3$ and the thermal de Broglie wavelength is $\lambda\equiv\sqrt{2\pi\hbar^{2}/(Mk_{B}T)}$.

Following the common practice, we define $\Delta b_{n,j} = b_{n,j} - b^{(1)}_{n,j}$, where the superscript ``(1)" denotes an ideal noninteracting system with the same fugacity. For a spin-polarized Fermi gas with coexisting $s$- and $p$-wave interactions, we may rewrite the thermodynamic potential as (up to the second order)
\begin{eqnarray}\label{omega}
\Omega(\mu_{\uparrow},\mu_{\downarrow})
&&=\Omega^{(1)} -\frac{1}{\beta}Q_{1}[z_{\uparrow}^{2}\Delta b^{(\uparrow\uparrow)}_{2,p} + z_{\uparrow}z_{\downarrow}\Delta b_{2,s}\nonumber\\
&&~~ + z_{\downarrow}^{2}\Delta b^{(\downarrow\downarrow)}_{2,p}],
\end{eqnarray}
where $\Omega^{(1)}$ is the thermodynamic potential for a noninteracting ideal Fermi gas with the same fugacities as $\Omega$. $\Delta b_{2,s}$ and $\Delta b^{(\sigma\sigma)}_{2,p}$ are, respectively, the second virial coefficients for the $s$- and $p$-wave scattering channels.

\subsection{Thermodynamic potential and number density}\label{3.1}

In an ideal spin-imbalanced Fermi gas, the thermodynamic potential for each spin species takes the form
\begin{eqnarray}\label{omega1}
\Omega^{(1)}(\mu_{\sigma}) = -\frac{1}{\beta}\frac{V}{\lambda^{3}}f_{5/2}(z_{\sigma}),
\end{eqnarray}
where the standard Fermi-Dirac integral is
\begin{eqnarray}\label{f}
f_{\upsilon}(z_{\sigma})=\frac{1}{\Gamma(\upsilon)}\int_{0}^{\infty}{\frac{x^{\upsilon-1}dx}{z_{\sigma}^{-1}e^{x}+1}}.
\end{eqnarray}
Here, $\Gamma(\upsilon)$ is the gamma function.

The total thermodynamical potential of an ideal spin-imbalanced Fermi gas is a direct sum of the thermodynamic potentials of each spin component \begin{eqnarray}
\Omega^{(1)}(\mu_{\uparrow},\mu_{\downarrow})&&=\Omega^{(1)}(\mu_{\uparrow})+\Omega^{(1)}(\mu_{\downarrow})\nonumber\\
&& = -\frac{1}{\beta}\frac{V}{\lambda^{3}}\left[f_{5/2}(z_{\uparrow})+f_{5/2}(z_{\downarrow})\right].
\end{eqnarray}

Accordingly, we can rewrite the thermodynamic potential of a strongly interacting polarized Fermi gas as (up to the second order)
\begin{eqnarray}\label{omegah}
\Omega
&&=-\frac{1}{\beta}\frac{V}{\lambda^{3}}\left[f_{5/2}(z_{\uparrow})+f_{5/2}(z_{\downarrow})+2z_{\uparrow}^{2}\Delta b^{(\uparrow\uparrow)}_{2,p}\right. \nonumber\\
&& \left.~~ + 2z_{\uparrow}z_{\downarrow}\Delta b_{2,s} + 2z_{\downarrow}^{2}\Delta b^{(\downarrow\downarrow)}_{2,p}\right].
\end{eqnarray}

Therefore, the particle number densities of spin-up and spin-down atoms are given by
\begin{eqnarray}\label{nh1}
n_{\uparrow}&&=-\frac{1}{V}\frac{\partial\Omega}{\partial\mu_{\uparrow}}\nonumber\\
&& = \frac{1}{\lambda^{3}}\left[f_{3/2}(z_{\uparrow})+2 z_{\uparrow}z_{\downarrow}\Delta b_{2,s}+4z_{\uparrow}^{2}\Delta b^{(\uparrow\uparrow)}_{2,p}\right],\\
\label{nh2}n_{\downarrow}&&=-\frac{1}{V}\frac{\partial\Omega}{\partial\mu_{\downarrow}}\nonumber\\
&& = \frac{1}{\lambda^{3}}\left[f_{3/2}(z_{\downarrow})+2 z_{\uparrow}z_{\downarrow}\Delta b_{2,s}+4z_{\downarrow}^{2}\Delta b^{(\downarrow\downarrow)}_{2,p}\right].
\end{eqnarray}

\subsection{Interaction energy}\label{3.2}

The total energy density can be written as
\begin{eqnarray}\label{e}
\epsilon=-\frac{1}{V}\left(\frac{\partial\ln\Xi}{\partial\beta}\right)_{z_{\sigma},V},
\end{eqnarray}
where the logarithm of the grand canonical partition function $\Xi$ is given by
\begin{eqnarray}\label{Xih}
\ln\Xi&&=-\beta\Omega\nonumber\\
&&=\frac{V}{\lambda^{3}}\left[f_{5/2}(z_{\uparrow})+f_{5/2}(z_{\downarrow})+2z_{\uparrow}^{2}\Delta b^{(\uparrow\uparrow)}_{2,p}\right. \nonumber\\
&& \left.~~ + 2z_{\uparrow}z_{\downarrow}\Delta b_{2,s} + 2z_{\downarrow}^{2}\Delta b^{(\downarrow\downarrow)}_{2,p}\right].
\end{eqnarray}

We can further separate the total energy density into the kinetic and the interaction part: $\epsilon=\epsilon_{kin}+\epsilon_{int}$. The kinetic energy density is \cite{Ho2004s,Ho2004p,Ho2012,Cui2012}
\begin{eqnarray}\label{ekin}
\epsilon_{kin}
&&=\frac{3 n_{\uparrow}k_{B}T}{2}\left[1+2^{-5/2}(n_{\uparrow}\lambda^{3})\right]\nonumber\\
&&~~+\frac{3 n_{\downarrow}k_{B}T}{2}\left[1+2^{-5/2}(n_{\downarrow}\lambda^{3})\right].
\end{eqnarray}
And the interaction energy density is given by
\begin{eqnarray}\label{eint}
\epsilon_{int}
&&=3k_{B}T\left[\left(-\Delta b_{2,s}+\frac{2}{3} T\Delta b^{'}_{2,s}\right)n_{\uparrow}(n_{\downarrow}\lambda^{3})\right. \nonumber\\
&& \left. ~~+\left(-\Delta b^{(\uparrow\uparrow)}_{2,p}+\frac{2}{3} T\Delta b^{'(\uparrow\uparrow)}_{2,p}\right)n_{\uparrow}(n_{\uparrow}\lambda^{3})\right. \nonumber\\
&& \left. ~~+\left(-\Delta b^{(\downarrow\downarrow)}_{2,p}+\frac{2}{3} T\Delta b^{'(\downarrow\downarrow)}_{2,p}\right)n_{\downarrow}(n_{\downarrow}\lambda^{3})\right],
\end{eqnarray}
where $\Delta b^{'}_{2,s}=d(\Delta b_{2,s})/dT$ and $\Delta b^{'(\sigma\sigma)}_{2,p}=d(\Delta b^{(\sigma\sigma)}_{2,p})/dT$.

For a spin-polarized Fermi gas with a fixed polarization $P=(n_{\uparrow}-n_{\downarrow})/n$, and a fixed total number density $n$, we can obtain  two different particle number densities $n_{\sigma}$. We then substitute the number densities into Eq.~(\ref{eint}) to get the corresponding interaction energy density with the second virial coefficients given in Sec.~\ref{3.4}.

\subsection{Contacts}\label{3.3}

The contacts can also be expressed in term of the grand thermodynamic potential $\Omega(\mu_{\uparrow},\mu_{\downarrow})$~\cite{Castin20121,Castin20122,contactHu2011,contactBlume2013}
\begin{eqnarray}
\label{Cs2}&&C_{s}=-\frac{4\pi M}{\hbar^{2}}\left(\frac{\partial \Omega}{\partial a_{s}^{-1}}\right)_{T,V,\mu_{\uparrow},\mu_{\downarrow}},\\
\label{CpV2}
&&C^{(\sigma\sigma)}_{p,\nu_{m}} =  -\frac{8\pi M}{\hbar^{2}} \left[\frac{\partial \Omega}{\partial (\nu_{m}^{(\sigma\sigma)})^{-1}}\right]_{T,V,\mu_{\uparrow},\mu_{\downarrow}},\\
\label{CpR2}
&&C^{(\sigma\sigma)}_{p,R_{m}} = -\frac{8\pi M}{\hbar^{2}} \left[\frac{\partial \Omega}{\partial (R_{m}^{(\sigma\sigma)})^{-1}}\right]_{T,V,\mu_{\uparrow},\mu_{\downarrow}}.
\end{eqnarray}
With these expressions, it is straightforward to numerically evaluate contacts and the associated universal thermodynamic properties once the second virial coefficients are known.

\subsection{Second virial coefficients}\label{3.4}

For a uniform system, the second virial coefficients can be expressed in terms of the phase shifts of the corresponding two-body scattering problem. For a spin-1/2 Fermi gas, it takes the form~\cite{Ho2004s,Ho2004p,liureview2013}
\begin{eqnarray}\label{Virial2}
\frac{\Delta b_{2}}{\sqrt{2}}
&&=\sum^{\infty}_{l=0}\sum^{l}_{m=-l}e^{-E_{B,l,m}/(k_{B}T)}\nonumber\\
&&~~+\sum^{\infty}_{l=0}\sum^{l}_{m=-l}\int_{0}^{\infty}e^{-\lambda^{2}k^{2}/(2\pi)}\frac{\partial\delta_{l,m}(k)}{\partial k}\frac{dk}{\pi},
\end{eqnarray}
where the summation is over all the two-body bound states (with the bounding energy $E_{B,l,m}$) and $\delta_{l,m}(k)$ is the phase shift of
the $l$th partial wave.

In the case of $s$-wave scattering~\cite{Rp-wave,Zhang2010p-wave,Peng2011p-wave}
\begin{eqnarray}\label{phaseshiftresonance}
k\text{cot}\delta_{0}(k)&&=-\frac{1}{a_{s}}+\frac{1}{2}r_{s,0}k^{2}+\cdot\cdot\cdot,
\end{eqnarray}
where $r_{s,0}$ is the effective range of interactions. Close to resonance, $|a_s|\gg r_{s,0}$, the second virial coefficient for $s$-wave scattering can then be written as~\cite{Ho2004s,Ho2004p}
\begin{eqnarray}\label{b2s}
\frac{\Delta b_{2,s}}{\sqrt{2}}&&=\Theta(a_{s})e^{-E_{B,s}/(k_{B}T)}
+\int_{0}^{\infty}e^{-\lambda^{2}k^{2}/(2\pi)}\frac{\partial\delta_{0}(k)}{\partial k}\frac{dk}{\pi}\nonumber\\
&&=\Theta(a_{s})e^{\tilde{x}^{2}}-\frac{\text{sign}(a_{s})}{2}\text{erfc}(\tilde{x})e^{\tilde{x}^{2}},
\end{eqnarray}
where $\tilde{x}=\sqrt{2 T_F/T}/(k_{F}|a_{s}|)$, $T_{F}$ is the Fermi temperature, $k_{F}$ is the Fermi wave vector, $\text{erfc}(\tilde{x})$ is the complementary error function, the Heaviside step function $\Theta(a_{s})$ in this paper is defined as $\Theta(a_{s})=0$ for $a_{s}<0$ and $\Theta(a_{s})=1$ for $a_{s}>0$,  the sign function $\text{sign}(a_{s})$ satisfies $\text{sign}(a_{s})=1$ with $a_{s}>0$, $\text{sign}(a_{s})=0$ with $a_{s}=0$, and $\text{sign}(a_{s})=-1$ with $a_{s}<0$, and $E_{B,s}=-\hbar^{2}/(Ma^2_s)$ is the $s$-wave two-body bound-state energy.

In the case of $p$-wave scattering~\cite{Rp-wave,Zhang2010p-wave,Peng2011p-wave}
\begin{eqnarray}\label{phaseshiftresonancep}
k^{3}\text{cot}\delta_{1,m}(k)=-\frac{1}{\nu_{m}}-\frac{k^{2}}{R_{m}}+\cdot\cdot\cdot,
\end{eqnarray}
where $\nu_m$ and $R_m$ are, respectively, the scattering volume and the effective range of the $p$-wave interaction with a magnetic number $m$.
The second virial coefficient in this case can be analytically given by~\cite{Yu2015}
\begin{eqnarray}\label{b2p}
\frac{\Delta b_{2,p}}{\sqrt{2}}
&&=\sum^{1}_{m=-1}\left[\Theta(\upsilon_{m})e^{-E_{B,m}/(k_{B}T)}\right. \nonumber\\
&& \left. ~~+\int_{0}^{\infty}e^{-\lambda^{2}k^{2}/(2\pi)}\frac{\partial\delta_{1,m}(k)}{\partial k}\frac{dk}{\pi}\right],
\end{eqnarray}
where $E_{B,m}=-\hbar^{2}R_{m}/(M\nu_{m})$ is the $p$-wave two-body bounding energy.

\section{Numerical results}\label{4}

In this section, we present our numerical results for the thermodynamic quantities based on the second virial coefficient calculations. To make connections with experiments, we consider $^{40}$K atoms close to the $198$G $p$-wave Feshbach resonance between the same hyperfine states $|F=9/2,m_F=-7/2\rangle$. As this $p$-wave resonance is close to a wide $s$-wave Feshbach resonance at $202.1$G between the hyperfine states $|F=9/2,m_F=-7/2\rangle$ and $|F=9/2,m_F=-9/2\rangle$, both $s$- and $p$-wave interactions exist in the system~\cite{exp2004s,exp2003p,review2010,exp2004p,Yu2015exp}.

Here, both the $s$- and $p$-wave scattering parameters are functions of the magnetic field.
For the $s$-wave scattering, the scattering length $a_s$ is given by~\cite{exp2004s,exp2003p,review2010}
\begin{eqnarray}
a_{s}(B)=a_{bg}\left(1-\frac{w}{B-B_{0,s}}\right),
\end{eqnarray}
where $B_{0,s}=202.1$G, $a_{bg}\simeq174a_{0}$, $a_{0}$ is the Bohr radius, and $w\simeq8.0$G.

For the $p$-wave scatterings, the scattering volume $\nu_m$ near the $p$-wave Feshbach resonance can be conveniently calculated using~\cite{review2010,exp2004p,Yu2015exp}
\begin{eqnarray}
\nu_{m=0}(B)&&=\nu_{z}(B)\nonumber\\
&&=a^{3}_{0}/(8.681\times10^{-5} - 8.29778\times10^{-7}\times B\nonumber\\
&&~~ + 1.97732\times10^{-9}\times B^{2}),\\
\nu_{m=\pm1}(B)&&=\nu_{xy}(B)\nonumber\\
&&=a^{3}_{0}/(7.83424\times10^{-5} - 7.456621\times10^{-7}\times B\nonumber\\
&&~~ + 1.76807\times10^{-9}\times B^{2}).
\end{eqnarray}
Here, there are two $p$-wave Feshbach resonances $B_{0,z}\simeq198.8$G and $B_{0,xy}\simeq198.3$G.

The interacting ranges for the $p$-wave interactions are~\cite{review2010,exp2004p,Yu2015exp}
\begin{eqnarray}
R_{m=0}(B)&&=R_{z}(B)\nonumber\\
&&=a_{0}/(1.64805 - 0.01523\times B\nonumber\\
&&~~ + 3.54471\times10^{-5}\times B^{2}),\\
R_{m=\pm1}(B)&&=R_{xy}(B)\nonumber\\
&&=a_{0}/(2.36792 - 0.02264\times B\nonumber\\
&&~~ + 5.45051\times10^{-5}\times B^{2}).
\end{eqnarray}

In the following, we will define $|F=9/2,m_F=-7/2\rangle\equiv|\uparrow\rangle$ and $|F=9/2,m_F=-9/2\rangle\equiv|\downarrow\rangle$. Therefore, the only $p$-wave interactions in the system are between two $|\uparrow\rangle$ states. For the numerical calculations, we take $1/k_{F}\sim100$nm, where the Fermi wave vector is $k_{F}=(3\pi^{2}n)^{1/3}$ with the total particle number density $n$.

\subsection{Interaction energy}\label{4.1}

With $\Delta b^{(\downarrow\downarrow)}_{2,p}=0$, the interaction energy density becomes
\begin{eqnarray}\label{eintK}
\epsilon_{int}
&&=3k_{B}T n_{\uparrow}\left[\left(-\Delta b_{2,s}+\frac{2}{3} T\Delta b^{'}_{2,s}\right)(n_{\downarrow}\lambda^{3})\right. \nonumber\\
&& \left. ~~+\left(-\Delta b^{(\uparrow\uparrow)}_{2,p}+\frac{2}{3} T\Delta b^{'(\uparrow\uparrow)}_{2,p}\right)(n_{\uparrow}\lambda^{3})\right].
\end{eqnarray}

\begin{figure}[tbp]
\includegraphics[width=8cm]{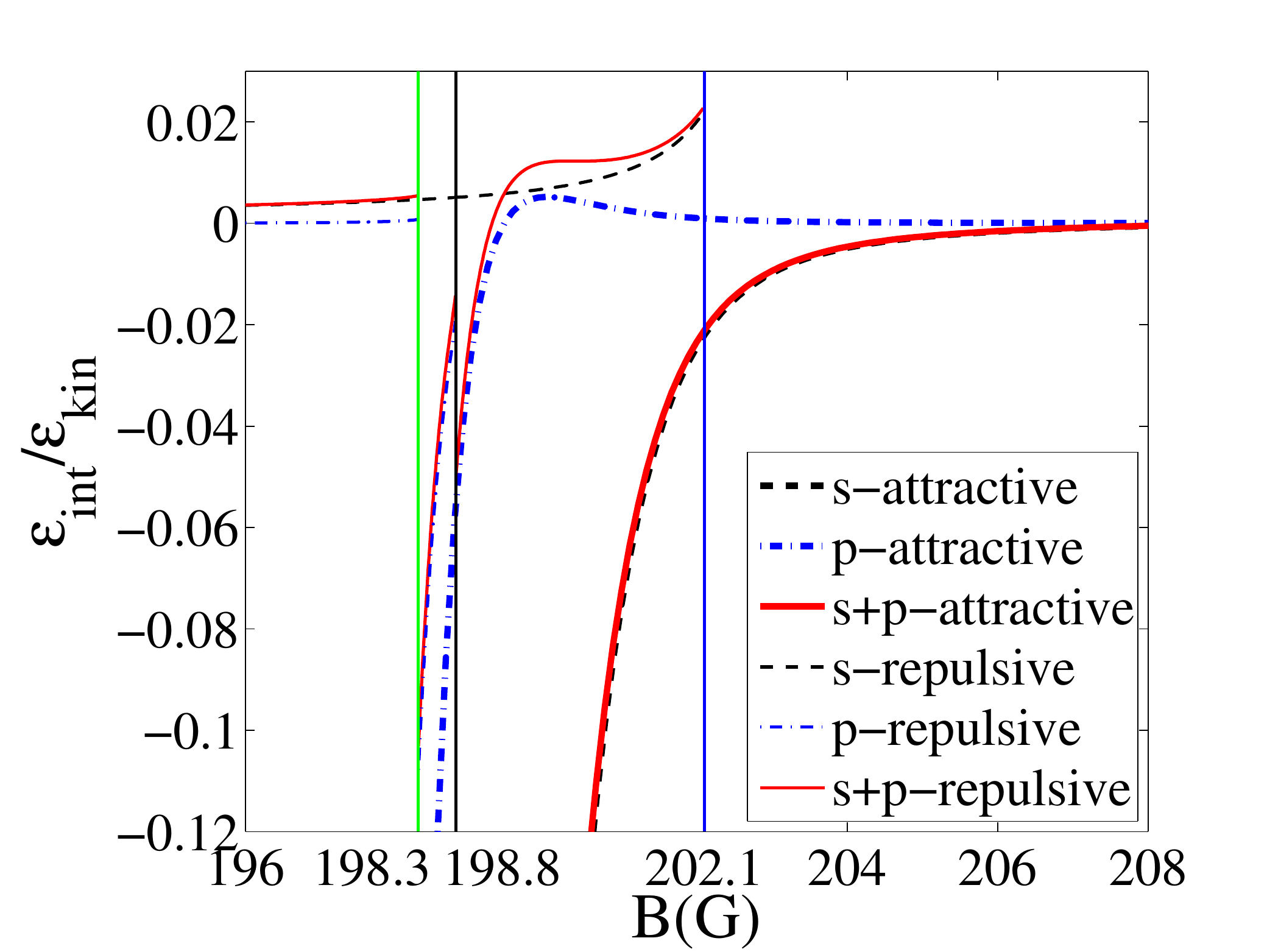}
\caption{(Color online) Interaction energy $\epsilon_{int}/\epsilon_{kin}$ of $^{40}$K atoms as a function of the magnetic field $B$ (in units of G) near $s$- and $p$-wave resonances for $T=5\mu K$~\cite{Zhang2010p-wave} with fixed spin polarization $P=0$. The blue solid line is for the $s$-wave Feshbach resonance at $B_{0,s}\simeq202.1$G, the black solid line denotes the $p$-wave Feshbach resonance with $m=0$ at $B_{0,z}\simeq198.8$G, and the green solid line denotes the $p$-wave Feshbach resonance with $m=\pm 1$ at $B_{0,xy}\simeq198.3$G.}
\label{interactionenergy}
\end{figure}

We show in Fig.~\ref{interactionenergy} the interaction energy density as a function of the magnetic field $B$ (in units of G) for $T=5\mu K$~\cite{Zhang2010p-wave} with a fixed spin polarization $P=0$. For comparison, we show the numerical results for a Fermi gas at the same temperature with only $s$-wave interactions (dashed black curve), with only $p$-wave interactions (dashed-dotted blue curve), and with both $s$- and $p$-wave interactions (solid red curve). We also show the interaction energy densities of the repulsive branch (thin curves), for which the bound-state contributions to the second virial coefficients are excluded in Eqs.~(\ref{b2s}) and (\ref{b2p}).

An interesting feature of Fig.~\ref{interactionenergy} is the behavior of the interaction energy of the repulsive branch across the $p$-wave Feshbach resonances. We consider  anisotropic $p$-wave interactions, in which the $p$-wave resonance for angular momentum $l=1, m=\pm1$ occurs $B_{0,xy}=198.3$G, and the $p$-wave resonance for angular momentum $l=1, m=0$ occurs at $B_{0,z}=198.8$G. In the case when only $p$-wave interactions are present, the interaction energy of the repulsive branch features an abrupt jump when the magnetic field is swept across the $198.3$G resonance. Physically, this is due to the coexistence of $p$-wave interactions with different magnetic angular momentum. For the case with coexisting $s$- and $p$-wave interactions, similar sudden changes of the repulsive-branch interaction energy show up at both $p$-wave resonances at $B=198.3$G and at $B=198.8$G, which originate from the coexistence of $s$- and $p$-wave interactions. Note that while the interaction energy in the repulsive branch approaches the pure $s$-wave value as the magnetic field decreases away from the $p$-wave resonance, throughout the $p$-wave resonance region, the interaction energy is quite different from that of the pure $s$-wave case. This behavior would provide a strong experimental signal for the existence of both scattering channels in the system.

\begin{figure*}[tbp]
\includegraphics[width=6cm]{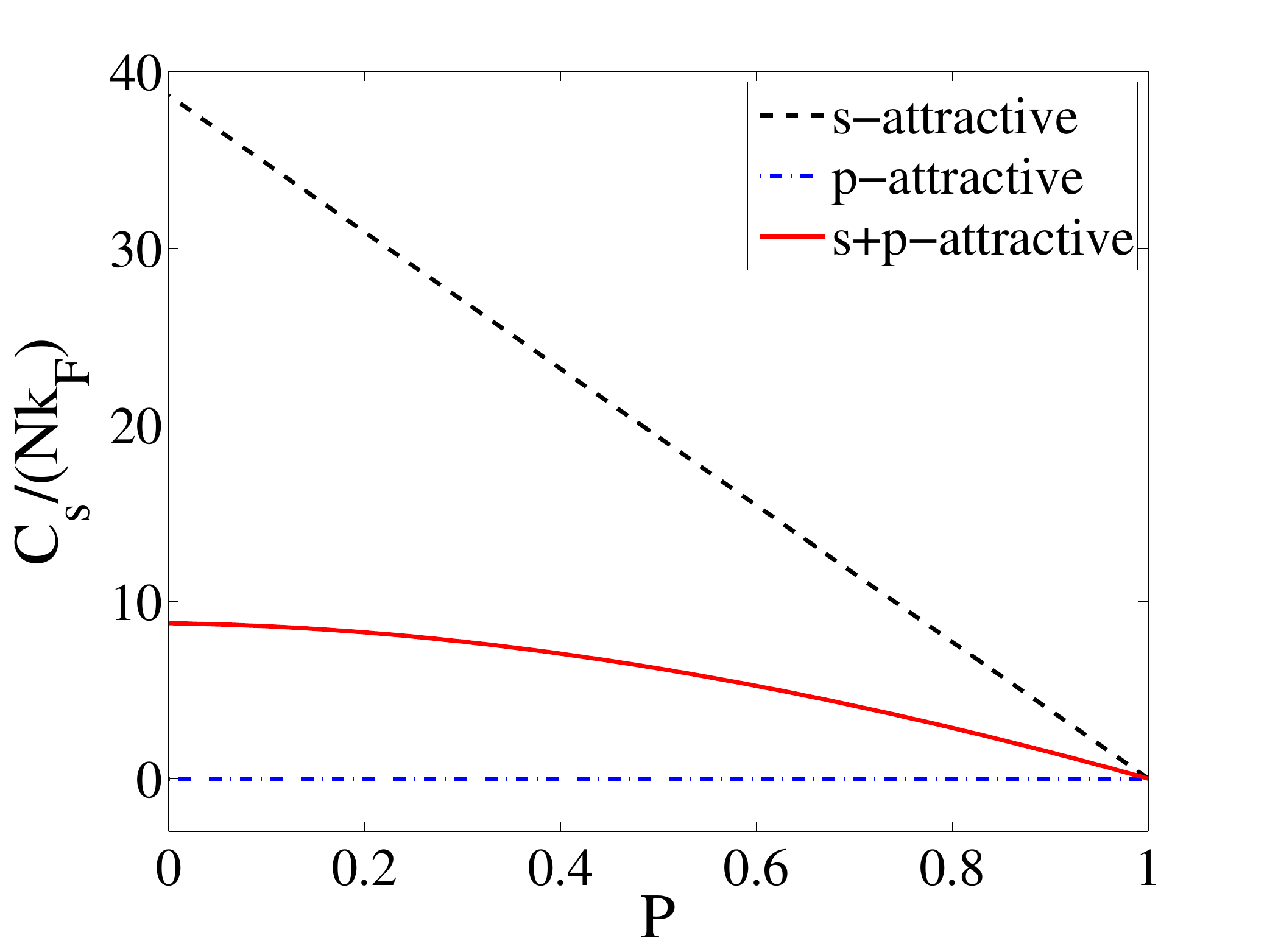}\includegraphics[width=6cm]{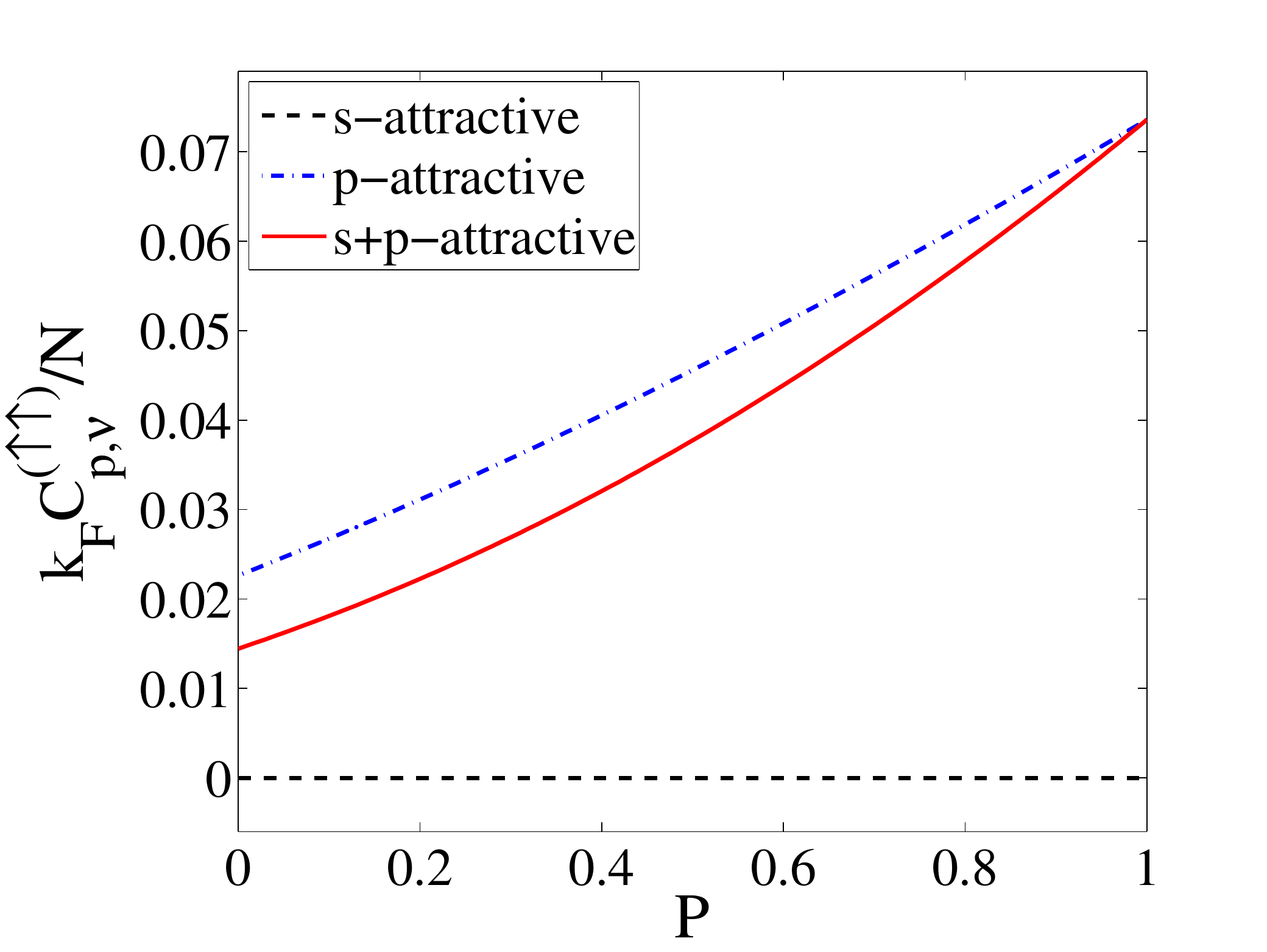}\includegraphics[width=6cm]{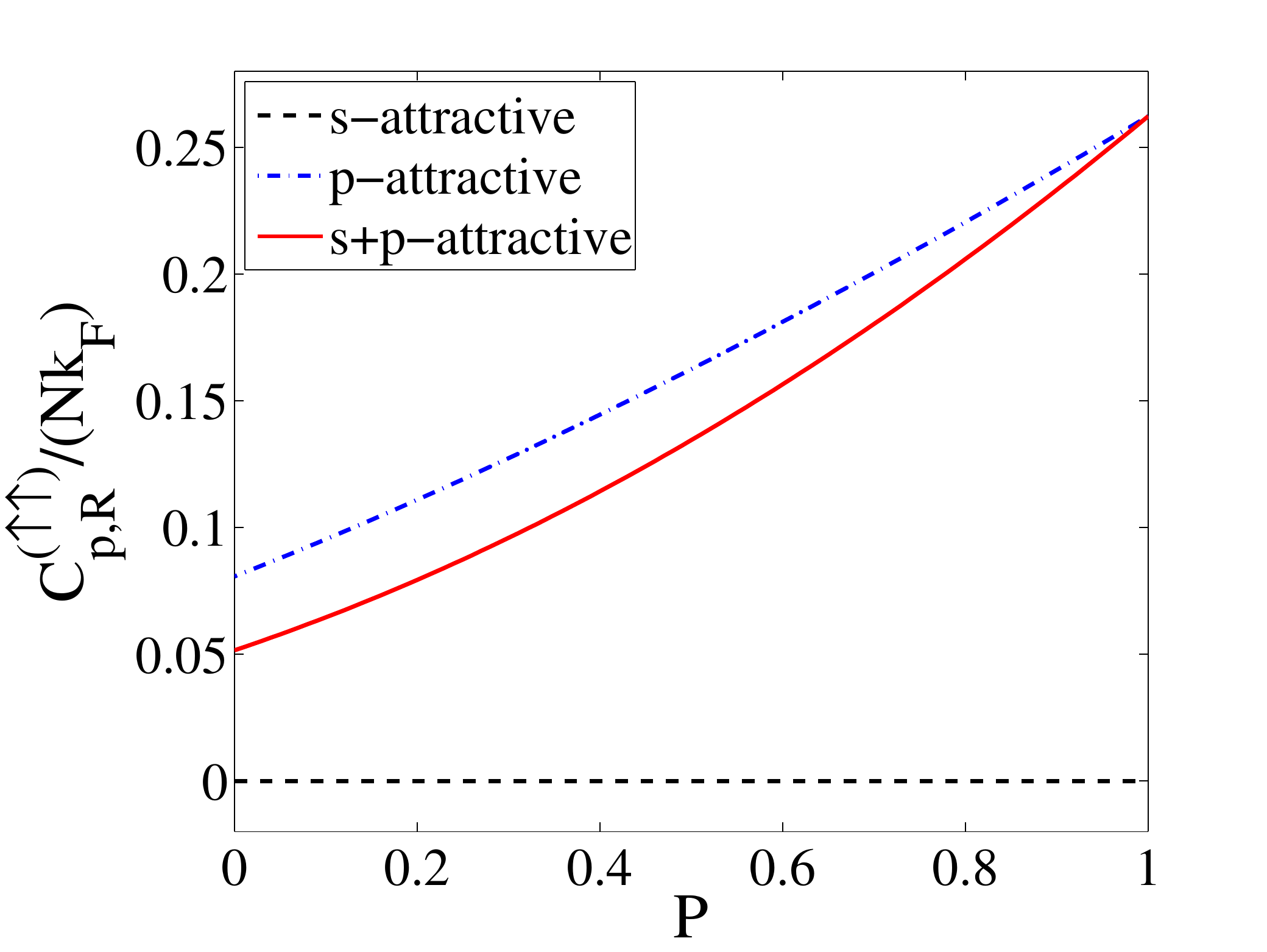}
\caption{(Color online) Contacts of a three-dimensional homogeneous strongly interacting Fermi gas
as functions of the spin polarization $P$ at a given temperature $T=10T_F$.
Here, we have chosen the magnetic field $B=199$G which is near the $p$-wave Feshbach resonance magnetic field for $^{40}$K atoms.}
\label{ContactP}
\end{figure*}

\subsection{Contacts}\label{4.2}

In order to calculate the contacts, we first expand the thermodynamic potential Eq.~(\ref{omegah}) with respect to a small fugacity $z_{\sigma}$:
\begin{eqnarray}\label{omegahcontact}
\Omega
&&\simeq-k_{B}T\frac{V}{\lambda^{3}}\left[z_{\uparrow}-\frac{z_{\uparrow}^{2}}{2^{5/2}}+z_{\downarrow}-\frac{z_{\downarrow}^{2}}{2^{5/2}}+2z_{\uparrow}^{2}\Delta b^{(\uparrow\uparrow)}_{2,p}\right. \nonumber\\
&& \left.~~ + 2z_{\uparrow}z_{\downarrow}\Delta b_{2,s}\right],
\end{eqnarray}
where we have taken $\Delta b^{(\downarrow\downarrow)}_{2,p}=0$.

The particle number densities Eqs.~(\ref{nh1}) and (\ref{nh2}) can also be expanded with respect to a small fugacity $z_{\sigma}$:
\begin{eqnarray}\label{nh12}
n_{\uparrow}&&\simeq \frac{1}{\lambda^{3}}\left[z_{\uparrow}-\frac{z_{\uparrow}^{2}}{2^{3/2}}+2 z_{\uparrow}z_{\downarrow}\Delta b_{2,s}+4z_{\uparrow}^{2}\Delta b^{(\uparrow\uparrow)}_{2,p}\right],\\
\label{nh22}n_{\downarrow}&&\simeq \frac{1}{\lambda^{3}}\left(z_{\downarrow}-\frac{z_{\downarrow}^{2}}{2^{5/2}}+2 z_{\uparrow}z_{\downarrow}\Delta b_{2,s}\right).
\end{eqnarray}

With Eqs.~(\ref{Cs2}), (\ref{CpV2}), (\ref{CpR2}), and Eq.~(\ref{omegahcontact}) the analytical expressions for the contacts become
\begin{eqnarray}
\label{Csup}C_{s}&&=-\frac{4\pi M}{\hbar^{2}}\left(\frac{\partial \Omega}{\partial a_{s}^{-1}}\right)_{T,V,\mu_{\uparrow},\mu_{\downarrow}}\nonumber\\
&&=\frac{8\pi M}{\hbar^{2}}k_{B}T\frac{V}{\lambda^{3}}\left(\frac{\partial\Delta b_{2,s}}{\partial a_{s}^{-1}}\right)_{T}z_{\uparrow}z_{\downarrow},\\
\label{CpVup}
C^{(\uparrow\uparrow)}_{p,\nu}&&=\sum_{m}C^{(\uparrow\uparrow)}_{p,\nu_{m}} =  -\frac{8\pi M}{\hbar^{2}} \sum_{m}\left[\frac{\partial \Omega}{\partial (\nu_{m}^{(\uparrow\uparrow)})^{-1}}\right]_{T,V,\mu_{\uparrow},\mu_{\downarrow}}\nonumber\\
&& = \frac{16\pi M}{\hbar^{2}} k_{B}T\frac{V}{\lambda^{3}}\sum_{m}\left[\frac{\partial\Delta b^{(\uparrow\uparrow)}_{2,p}}{\partial(\nu_{m}^{(\uparrow\uparrow)})^{-1}}\right]_{T}z_{\uparrow}^{2},\\
\label{CpRup}
C^{(\uparrow\uparrow)}_{p,R}&&=\sum_{m}C^{(\uparrow\uparrow)}_{p,R_{m}} = -\frac{8\pi M}{\hbar^{2}} \sum_{m}\left[\frac{\partial \Omega}{\partial (R_{m}^{(\uparrow\uparrow)})^{-1}}\right]_{T,V,\mu_{\uparrow},\mu_{\downarrow}}\nonumber\\
&& = \frac{16\pi M}{\hbar^{2}} k_{B}T\frac{V}{\lambda^{3}}\sum_{m}\left[\frac{\partial\Delta b^{(\uparrow\uparrow)}_{2,p}}{\partial (R_{m}^{(\uparrow\uparrow)})^{-1}}\right]_{T}z_{\uparrow}^{2}.
\end{eqnarray}

Numerically, we calculate the two fugacities by solving the coupled number equations Eqs.~(\ref{nh12}) and (\ref{nh22}), and then substitute the fugacities into Eqs.~(\ref{Csup}), (\ref{CpVup}), and (\ref{CpRup}) to get the corresponding contacts.
For convenience, in the following discussions, we only calculate the attractive branch of the contacts.

Figure~\ref{ContactP} shows the contacts as functions of spin polarization $P$ for a homogeneous Fermi gas up to the second-order virial expansion with $T=10T_F$ and $B=199$G (solid red line).
Here, the Fermi temperature is given by $T_{F}=(3\pi^{2}n)^{2/3}\hbar^2/(2M)/k_{B}$.
For comparison, we also show the numerical results for system with only $s$-wave interactions (dashed black line), and with only $p$-wave interactions (dashed-dotted blue line). Note that $C_s=0$ ($C^{(\uparrow\uparrow)}_{p,\nu/R}=0$) for systems with only $p$-wave ($s$-wave) interactions.

\begin{figure*}[tbp]
\includegraphics[width=6cm]{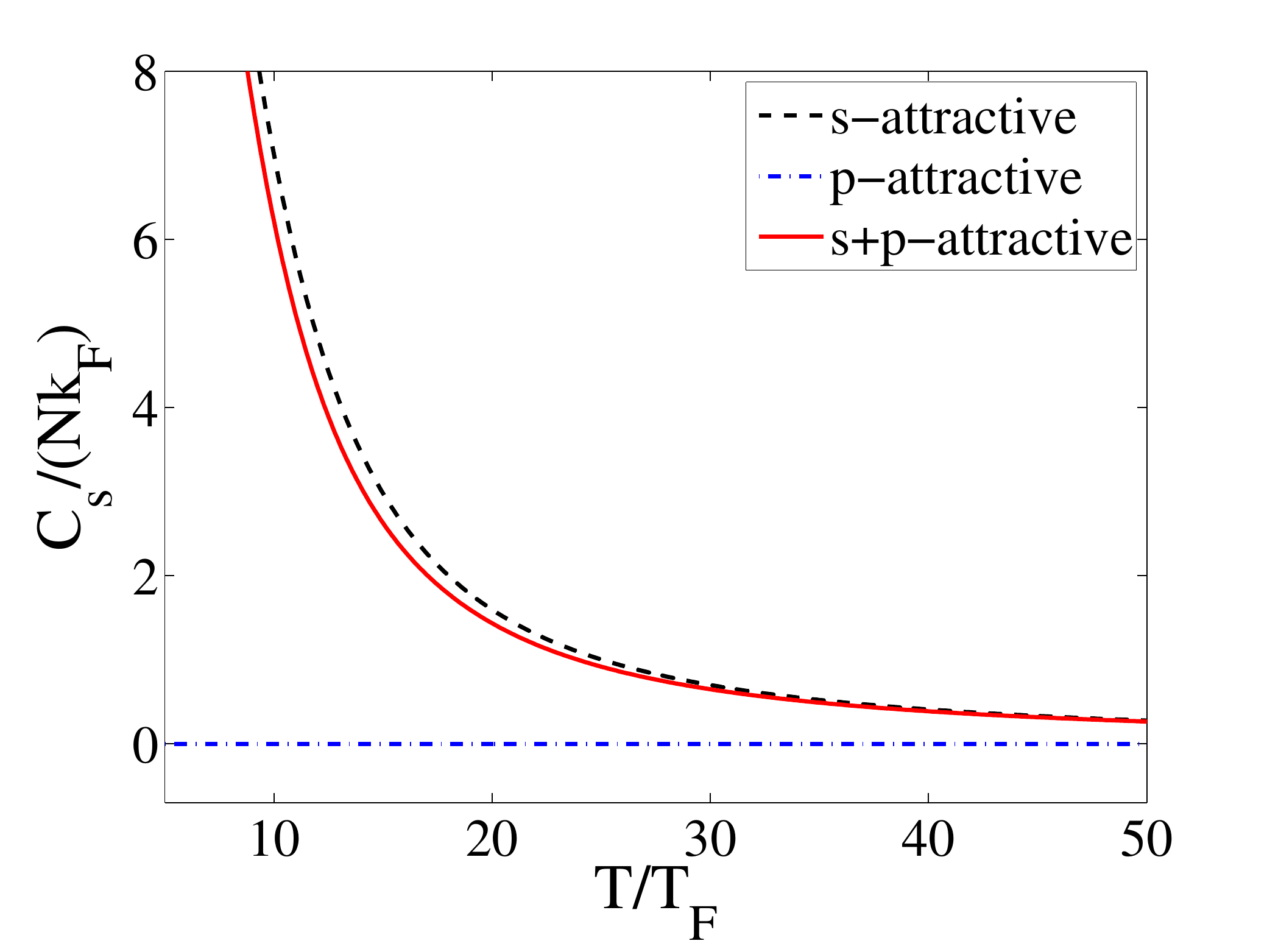}\includegraphics[width=6cm]{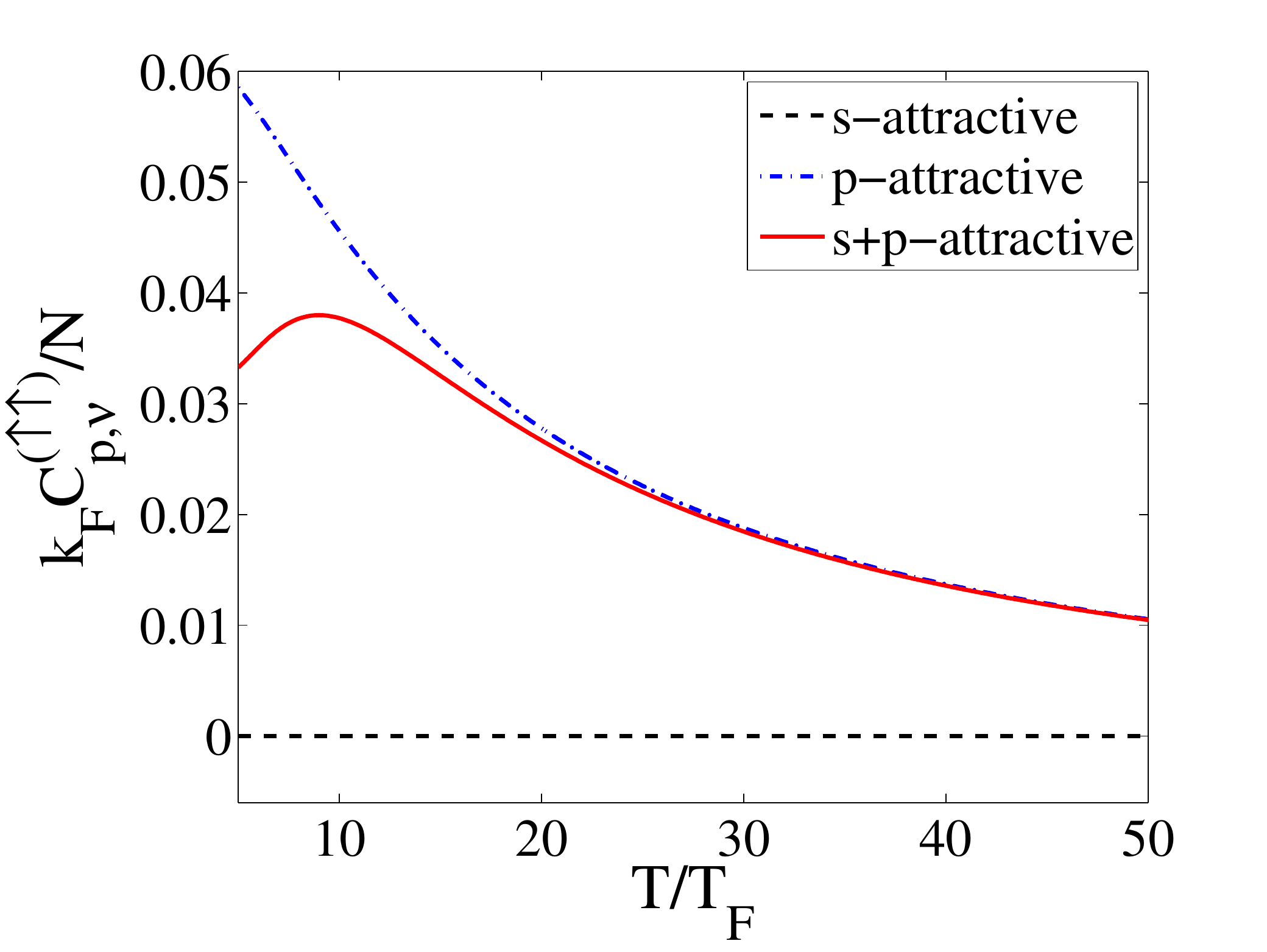}\includegraphics[width=6cm]{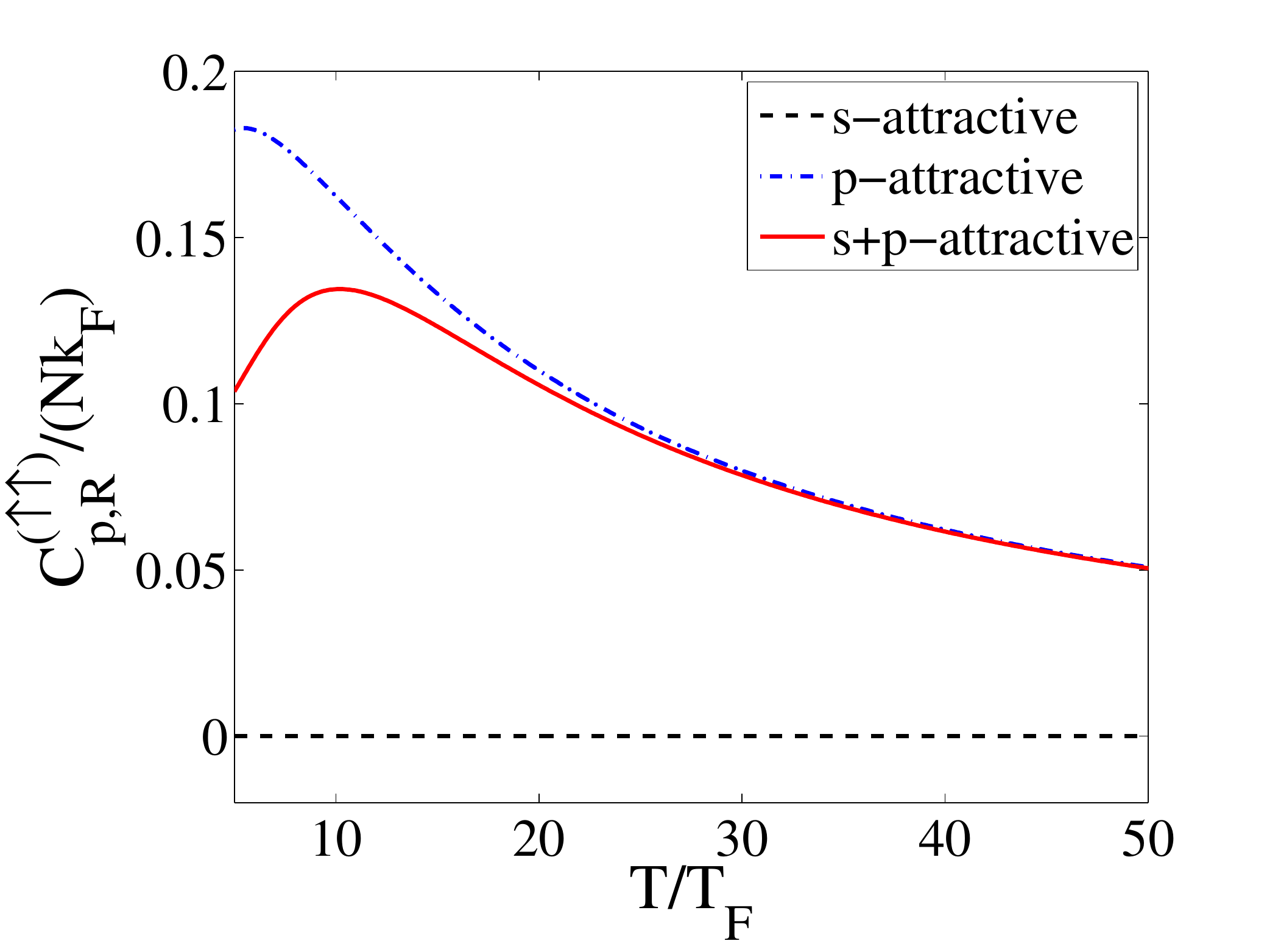}
\caption{(Color online) Temperature dependence of the contacts
of a three-dimensional homogeneous strongly interacting Fermi gas with the spin polarization $P=0.5$
and magnetic field $B=199$G.
The line styles are similar to Fig.~\ref{ContactP}.}
\label{ContactT}
\end{figure*}

In Fig.~\ref{ContactT}, we show the temperature dependence of various contacts with a polarization $P=0.5$, and a magnetic field $B=199$G.
Evidently, at lower temperatures, $C_s$ and $C^{(\uparrow\uparrow)}_{p,\nu/R}$ for the case with coexisting $s$- and $p$-wave interactions can be significantly different from those with purely $s$- or $p$-wave interactions. At high temperatures, as thermal effects dominate over interaction effects, all the contacts approach zero.

\begin{figure*}[tbp]
\includegraphics[width=6cm]{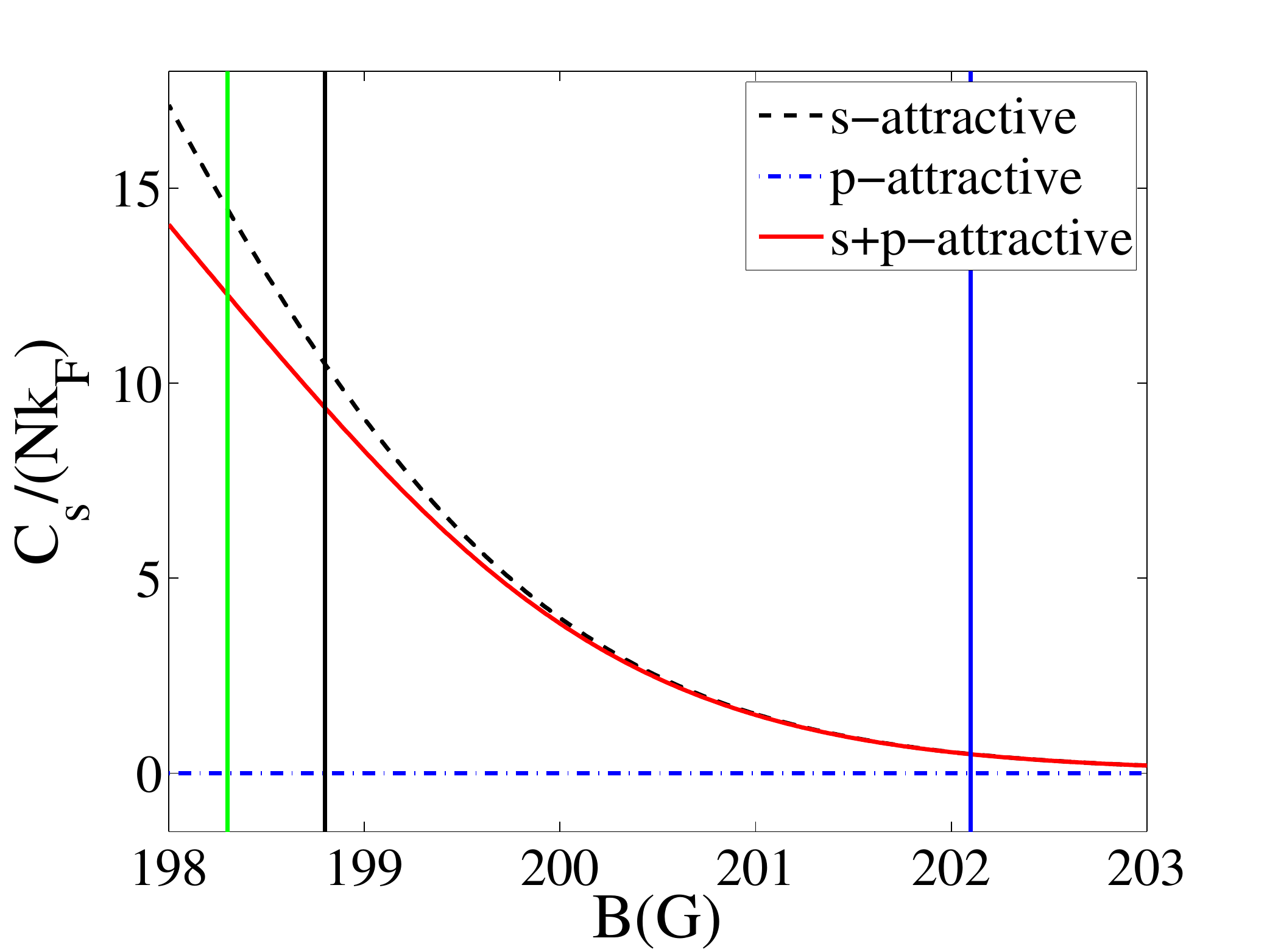}\includegraphics[width=6cm]{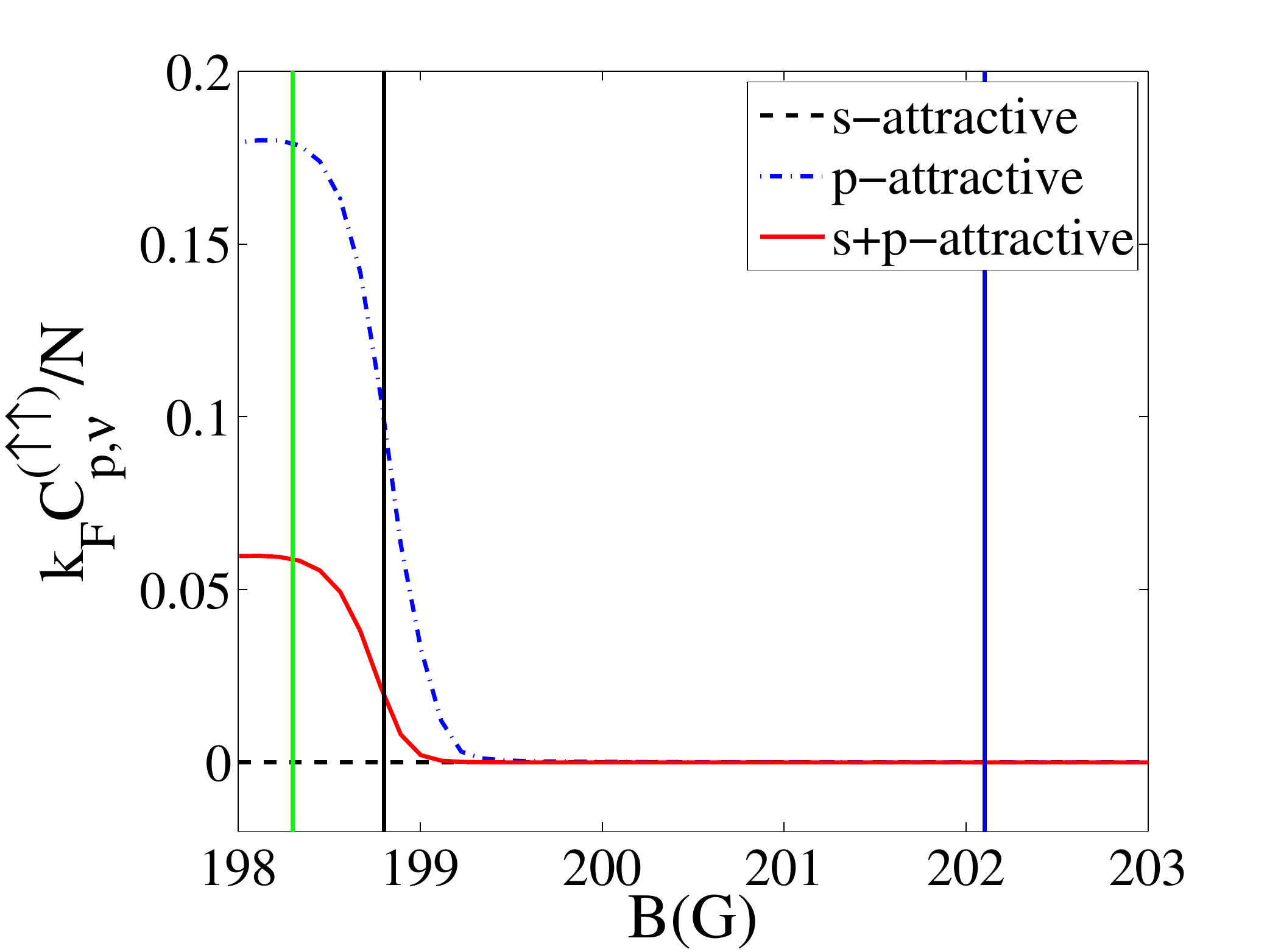}\includegraphics[width=6cm]{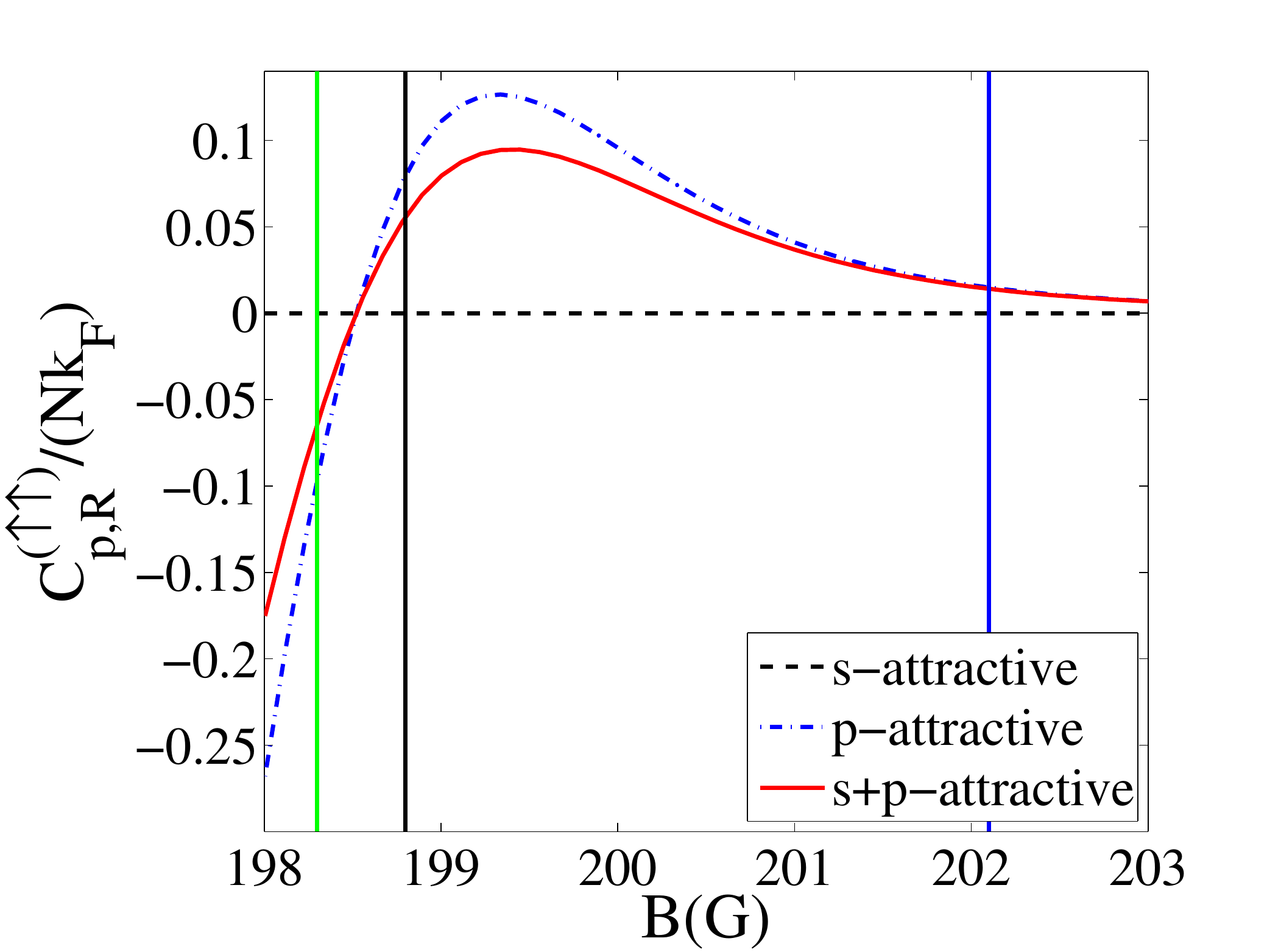}
\caption{(Color online) Contacts as functions of the magnetic field magnitude changing from $B=198$G to $B=203$G along the two $p$-wave resonances to the $s$-wave resonance with the spin polarization $P=0.2$ at a given temperature $T=10T_F$.
Here, we show the $p$-wave Feshbach resonance magnetic field $B_{0,xy}=198.3$G by a green solid line,
another $p$-wave Feshbach resonance $B_{0,z}=198.8$G by a black solid line,
and the $s$-wave Feshbach resonance $B_{0}=202.1$G by a blue solid line.
The other line styles are the same as in Fig.~\ref{ContactP}.}
\label{ContactB}
\end{figure*}

Finally, in Fig.~\ref{ContactB}, we show the dependence of contacts on the magnetic field with a polarization $P=0.2$ at a given temperature $T=10T_F$. Apparently, the interplay of $s$- and $p$-wave interactions is most significant near the $p$-wave resonances. We also note an interesting sign change of $C^{(\uparrow\uparrow)}_{p,R}$ as the magnetic field is swept between two $p$-wave resonances.

From Figs.~\ref{ContactP},\ref{ContactT}, and \ref{ContactB}, it appears that, in the presence of mixed $s$- and $p$-wave interactions, despite the apparent decoupling of different scattering channels in the universal relations, both the $s$- and the $p$-wave contacts are different from those with a single scattering channel. This clearly demonstrates the interesting interplay between $s$- and $p$-wave interactions on the many-body level. Physically, as the spin-up atoms are involved simultaneously in both scattering channels, the inclusion of any one of the two scattering channels on top the other would lead to a spin-dependent shift of the chemical potentials for the many-body system. This amounts to a change of the many-body environment, although the microscopic two-body physics in each channel remains decoupled and unchanged. In terms of numerical calculation, this interplay is reflected in the solutions of the coupled number equations Eqs.~(\ref{nh12}) and (\ref{nh22}).

\subsection{Momentum distributions}\label{4.3}

Experimentally, contacts can be determined by measuring the asymptotic behavior of the momentum distribution at large momenta. In a system with coexisting $s$- and $p$-wave interactions, this high-momentum tail would also behave differently.

To see this, we define the single-body density matrix in the asymptotic regime $r_0\ll r\ll k_{\rm F}^{-1}$ with the interaction potential range $r_0$ and $r=|\vec{r}|=|\vec{r}_1-\vec{r}_2|$:
\begin{equation}
\rho_1^{(\sigma)}(\vec{r}_1,\vec{r}_2)=\sum_{\eta,\gamma}\int d^3 \vec{r}_3 \langle\psi_{\sigma}^\dagger(\vec{r}_1)\psi_{\eta}^\dagger(\vec{r}_3)\psi_{\gamma}(\vec{r}_3)\psi_{\sigma}(\vec{r}_2)\rangle.
\end{equation}
Similarly to the approach in Sec.~\ref{2}, it is straightforward to derive
\begin{widetext}
\begin{eqnarray}\label{rho1up}
\rho_1^{(\uparrow)}(\vec{r}_1,\vec{r}_2)&&= \frac{1}{V}\int d^3 \vec{r}_{3} \sum_{\vec{P}_{c},j,m}n^{(\uparrow)}_{\vec{P}_{c},j,1,m}e^{i\vec{P}_{c}\cdot(\vec{r}_2-\vec{r}_1)/2}\nonumber\\
&&~~ \times \int dk\int dk'~b_{j,m,k}^{(\uparrow\uparrow)}b_{j,m,k'}^{(\uparrow\uparrow)}Y_{1,m}^{*}(\theta,\varphi)Y_{1,m}(\theta',\varphi')\left[ \frac{1}{|\vec{r}_1-\vec{r}_3|^2|\vec{r}_3-\vec{r}_2|^2} + \frac{k^2}{2|\vec{r}_1-\vec{r}_3|^2} + \frac{k'^2}{2|\vec{r}_3-\vec{r}_2|^2}\right]\nonumber\\
&&~~ + \frac{1}{V}\int d^3 \vec{r}_3 \sum_{\vec{P}_{c},j}n^{(\uparrow\downarrow)}_{\vec{P}_{c},j,0,0}e^{i\vec{P}_{c}\cdot(\vec{r}_2-\vec{r}_1)/2}\int dk~a_{j,k}\int dk'~a_{j,k'}\frac{Y_{0,0}^{*}(\theta,\varphi)}{|\vec{r}_1-\vec{r}_3|}\frac{Y_{0,0}(\theta',\varphi')}{|\vec{r}_3-\vec{r}_2|},\\
\label{rho1down}
\rho_1^{(\downarrow)}(\vec{r}_1,\vec{r}_2)&&= \frac{1}{V}\int d^3 \vec{r}_3 \sum_{\vec{P}_{c},j}n^{(\uparrow\downarrow)}_{\vec{P}_{c},j,0,0}e^{i\vec{P}_{c}\cdot(\vec{r}_2-\vec{r}_1)/2}\int dk~a_{j,k}\int dk'~a_{j,k'}\frac{Y_{0,0}^{*}(\theta,\varphi)}{|\vec{r}_1-\vec{r}_3|}\frac{Y_{0,0}(\theta',\varphi')}{|\vec{r}_3-\vec{r}_2|}.
\end{eqnarray}
\end{widetext}
Note that we have already assumed that no $p$-wave interactions exist between spin-$|\downarrow\rangle$ atoms.

Fourier-transforming Eqs.~(\ref{rho1up}) and (\ref{rho1down}), we obtain the following asymptotic behavior of $n_{\sigma}(\vec{k})$ for $k_F\ll k\ll 1/r_0$:
\begin{eqnarray}\label{nhup}
n_{\uparrow}(\vec{k}) && =  \sum_{m}Y_{1,m}^{*}(\hat{k})Y_{1,m}(\hat{k})\nonumber\\
&& ~~\times\left[\frac{4\pi C^{(\uparrow\uparrow)}_{p,\nu_{m}}}{Vk^{2}} + \frac{8\pi C^{(\uparrow\uparrow)}_{p,R_{m}}}{Vk^{4}}\right. \nonumber\\
&& \left.~~ - \frac{8\pi^2}{3}\frac{\sum_{\vec{P}_{c},j}n^{(\uparrow)}_{\vec{P}_{c},j,1,m}(\int dk'~b_{j,m,k'}^{(\uparrow\uparrow)})^{2}P^{2}_{c}}{Vk^{4}}\right]\nonumber\\
&& ~~ + \pi\frac{\sum_{\vec{P}_{c},j,m}n^{(\uparrow)}_{\vec{P}_{c},j,1,m}(\int dk'~b_{j,m,k'}^{(\uparrow\uparrow)})^{2}P^{2}_{c}}{Vk^{4}}\nonumber\\
&& ~~ + \frac{C_{s}}{Vk^{4}},\\
\label{nhdown}
n_{\downarrow}(\vec{k}) && = \frac{C_{s}}{Vk^{4}}.
\end{eqnarray}
Here, we assume that the distribution of $\vec{P}_{c}$ is isotropic, so we have $\langle P_{c,i} \rangle=0$ and $\langle P_{c,i}^{2}\rangle=\langle\vec{P}_{c}^{2}\rangle/3~(i=x,y,z)$.
Equations.~(\ref{nhup}) and (\ref{nhdown}) show that the momentum distributions are highly asymmetric in spin. Especially, the momentum distribution of the spin-$\uparrow$ component exhibits a quite nontrivial high-momentum tail, which is due to the coexistence of $s$- and $p$-wave interactions.

\subsection{Radio-frequency spectroscopies}\label{4.4}

Another experimental tool for the detection of contacts is the r.f. spectroscopy, in which universal scalings in the high frequency tail exist and are governed by contacts~\cite{Yu2015,rf2010s-wave1,rf2010s-wave2}. The r.f. coupling Hamiltonian $\mathcal{H}_{rf}=\hbar\Omega_{rf}\int d^{3}\vec{r}~\psi_{f}^{\dag}(\vec{r})\psi(\vec{r})$ transfers fermions into a third spin state $|f\rangle$, where $\Omega_{rf}$ is the radio-frequency Rabi frequency determined by the strength of the radio-frequency signal with frequency $\omega$.
Furthermore, the transfer rate can be evaluated as
\begin{align}
\Gamma_{rf}(\omega)=\frac{\pi}{\hbar^2}\sum_{i,f}\rho_{i}|\langle f|\mathcal{H}_{rf}|i\rangle|^{2}\delta[\hbar\omega-(E_{f}-E_{i})],
\end{align}
where $i$, $f$ label the initial and final states, $\rho_{i}$ denotes the initial state distribution, and $E_{i}$, $E_{f}$ denote the initial and final energies. In the region $E_R\equiv\hbar^2/(MR^2)\gg\hbar\omega\gg E_F$ ($E_F=\hbar^2k_F^2/(2M)$ is the Fermi energy), the asymptotic radio-frequency transition rates for the two spin species become
\begin{eqnarray}\label{rfup}
\Gamma_{rf,\uparrow}(\omega)&&=\pi\Omega_{rf}^{2}\sum_{i,f}\int d^{3}\vec{r}~\rho_1^{(\uparrow)}(\vec{r})\delta\left[\hbar\omega-\left(E_f-E_i\right)\right]\nonumber\\
&& =  \frac{M\Omega_{rf}^{2}}{2\pi\hbar}\sum_{m=-1}^{1}\left[\frac{C^{(\uparrow\uparrow)}_{p,\nu_{m}}}{(M\omega/\hbar)^{1/2}} + \frac{3C^{(\uparrow\uparrow)}_{p,R_{m}}}{2(M\omega/\hbar)^{3/2}}\right]\nonumber\\
&&~~ + \frac{M\Omega_{rf}^{2}}{4\pi\hbar}\frac{C_{s}}{(M\omega/\hbar)^{3/2}},\\
\label{rfdown}
\Gamma_{rf,\downarrow}(\omega)&&=\frac{M\Omega_{rf}^{2}}{4\pi\hbar}\frac{C_{s}}{(M\omega/\hbar)^{3/2}}.
\end{eqnarray}

\begin{figure}[tbp]
\includegraphics[width=8cm]{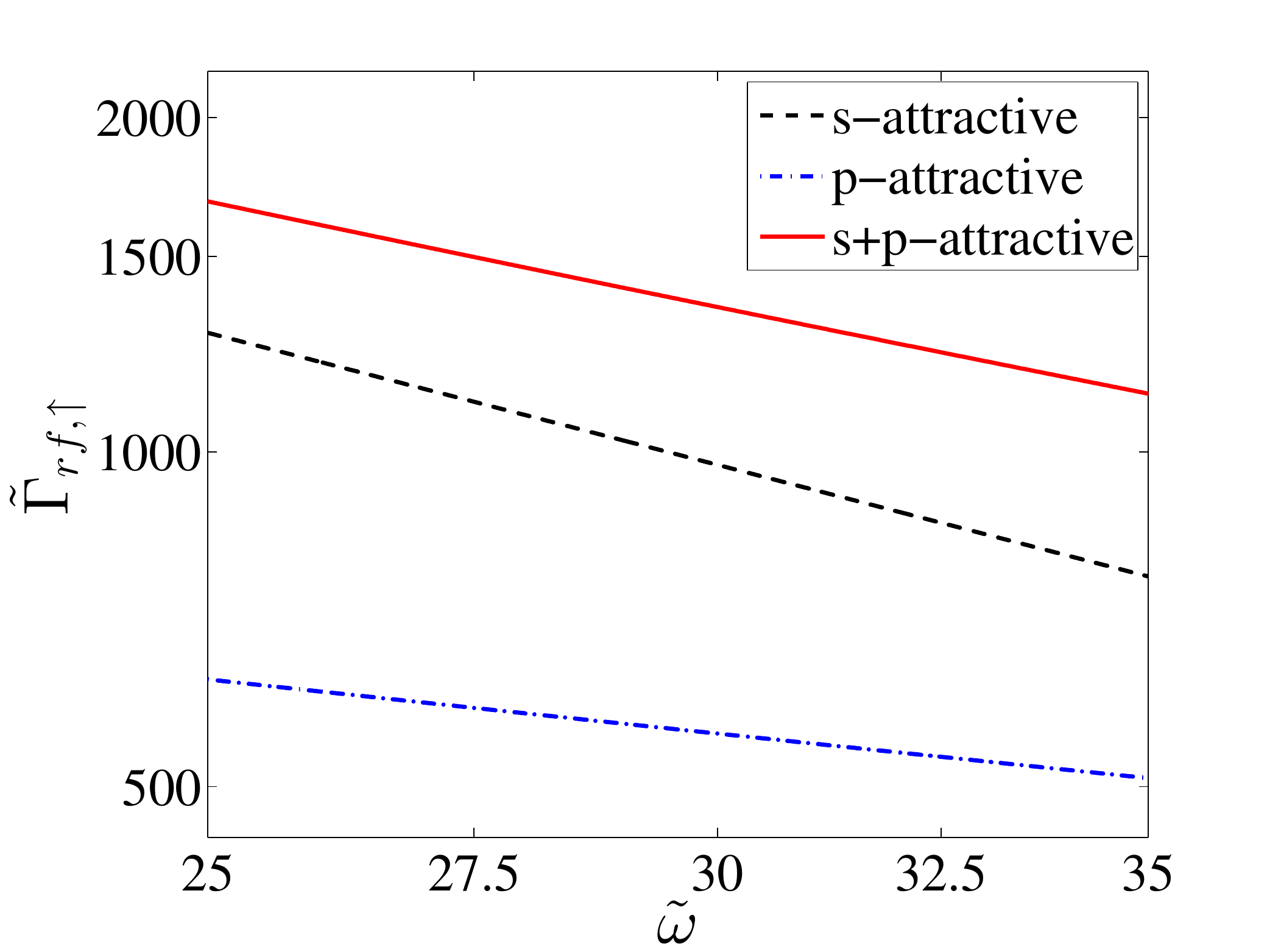}
\caption{(Color online) Reduced radio-frequency transition rate $\tilde{\Gamma}_{rf,\uparrow}=\Gamma_{rf,\uparrow}(\omega)\hbar/(VM\Omega_{rf}^{2})$ with spin-$\uparrow$ as a function of the reduced frequency $\tilde{\omega}=M\omega/(\hbar k_{F}^{2})$ at magnetic field $B=199$G for high frequency at temperature $T=10T_F$ with spin polarization $P=0.6$.
The line styles are the same as in Fig.~\ref{ContactP}.}
\label{rf}
\end{figure}

Again, Eqs.~(\ref{rfup}) and (\ref{rfdown}) show that the radio-frequency transition rates are highly asymmetric in spin. In Fig.~\ref{rf}, we show the reduced radio-frequency transition rate $\tilde{\Gamma}_{rf,\uparrow}=\Gamma_{rf,\uparrow}(\omega)\hbar/(VM\Omega_{rf}^{2})$ with spin-$\uparrow$ as a function of the reduced frequency $\tilde{\omega}=M\omega/(\hbar k_{F}^{2})$ at magnetic field $B=199$G for the temperature $T=10T_F$ with spin polarization $P=0.6$. Like the high-momentum tail of the momentum distribution, the high-frequency tail of the r.f. spectroscopy demonstrates the effects of coexistence of $s$- and $p$-wave interactions.

\section{Summary}\label{5}

In this work, we study the universal relations and normal-phase thermodynamic properties of a two-component Fermi gas with coexisting $s$- and $p$-wave interactions. Similarly to the purely $s$-wave or $p$-wave case, we start from the two-body density matrix and derive the adiabatic relations, through which a set of contacts can be defined. Due to the orthogonality of the two-body wave functions in different scattering channels, contacts from different scattering channels are decoupled. We may therefore identify $s$-wave contact and $p$-wave contacts based on the adiabatic relations. We then numerically evaluate the interaction energy and contacts using second-order virial expansions. For typical experimental parameters corresponding to $^{40}$K atoms, we find that when the interplay of $s$- and $p$-wave interactions is properly taken into account, all contacts behave differently
from those with a single scattering channel. We characterize in detail the dependence of contacts on quantities such as spin polarization, temperature, and magnetic field. An interesting finding is that the interaction energy of the repulsive branch features abrupt changes as the $p$-wave resonances are crossed. Our results are closely related to the ongoing efforts of realizing and probing strongly interacting quantum gases with high partial-wave interactions. As the interplay of different partial-wave interactions can have important effects on the many-body properties, our work has interesting implications for experiments exploring the $198$G $p$-wave resonance in $^{40}$K atoms, where interactions in different partial-wave scattering channels naturally coexist.

\section*{Acknowledgements}
We thank Zhenhua Yu, Shizhong Zhang, Shao-Liang Zhang, Su Wang, and Jian-Song Pan for helpful discussions.
This work is supported by the National Key R\&D Program (Grant No. 2016YFA0301700), the National Natural Science Foundation of China (Grants No. 11374177, No. 11374283, No. 11404106, No. 11522545, No. 11534014, No. 11622436), and the programs of the Chinese Academy of Sciences. F.Q. acknowledges support from the project funded by the China Postdoctoral Science Foundation (Grant No. 2016M602011). W.Y. acknowledges support from the ``Strategic Priority Research Program (B)'' of the Chinese Academy of Sciences, Grant No. XDB01030200.

\end{CJK*}
\end{document}